# Tuning Bound States of Symmetry-Breaking Vortices via Unidirectional Charge Density Wave in a Transition-Metal Dichalcogenide Superconductor

Hao Zhang[1,2,#], Hui Chen[1,2,3,#,*], Zichen Huang[1,2,#], Zi-Ang Wang[1,2], Senhao Lv[1,2], Guoyu Xian[1,2], Hui Guo[1,2,3,*], Haitao Yang[1,2,3] and Hong-Jun Gao[1,2,3,*]

[1] Beijing National Center for Condensed Matter Physics and Institute of Physics, Chinese Academy of Sciences, Beijing 100190, PR China

[2] School of Physical Sciences, University of Chinese Academy of Sciences, Beijing 100190, PR China

[3] Hefei National Laboratory, 230088 Hefei, Anhui, PR China

[#]These authors contributed equally to this work

[*]Correspondence to: hjgao@iphy.ac.cn, hchenn04@iphy.ac.cn, guohui@iphy.ac.cn

# ABSTRACT

The interplay between charge density wave (CDW) and superconducting vortex bound states are crucial for fundamental physics of superconductivity and advancing quantum nanotechnologies. However, the CDW-mediated modulation of vortex bound states, which opens up a new platform for vortex engineering, remains unexplored. Here, we report spatially anisotropic vortex states modulated by the unidirectional CDWs in a transition-metal dichalcogenide superconductor 1T″-NbTe$_2$ using ultra-low-temperature scanning tunneling microscopy/spectroscopy. The stripe-like 3×1×3 CDW order exhibits a robust three-dimensional character across step edges and coexists with superconductivity below a critical temperature of 0.4 K. Under out-of-plane magnetic fields, we observe elliptical vortices whose elongation aligns with the CDW stripes, indicating strong coupling between vortex morphology and underlying electronic order. Remarkably, CDW domain boundaries induce abrupt changes in vortex orientation and vortex bound states, enabling controllable vortex states across CDW nanodomains. These findings establish a new pathway for manipulating superconducting vortex bound states via CDW coupling.

**KEYWORDS:** *Transition metal dichalcogenide, symmetry breaking, vortex bound state, unidirectional charge density wave, domain wall*

Studies of superconducting vortices arising from quantized magnetic flux penetrating type-II superconductors provide critical insights into the fundamental properties of superconducting states[1,2]. As the smallest magnetic entities in these materials, vortices play a key role in the development of superconducting electronic devices[3,4]. One of the most promising frontiers is topological quantum computation. In particular, the vortex cores of intrinsic or effective topological superconductors are predicted to host zero-energy Majorana bound states[5–8]. The ability to accurately and rapidly manipulate individual vortex core states is the key prerequisite for realizing topological qubits based on these bound states[6]. Recent strategies for vortex control include the use of local thermal effects[9], magnetic force[10] and lattice defects[11]. In addition to positional control, understanding the spatial symmetry of vortex core states provides further insights into the superconducting pairing symmetries [12] and quantized magnetic flux[13,14].

Experimentally, significant progress has been made in obtaining the symmetry-breaking vortex structures in candidate topological superconductors. While vortices in most conventional superconductors are typically cylindrical with circular cross-sections[15], sixfold starshaped vortex structure has been observed in some superconductors with hexagonal lattice including 2H-NbSe$_2$[16], La[17] and kagome superconductors[18]. In iron-based superconductors, vortices are usually round or fourfold symmetric[19], though various distinct patterns have been reported in specific cases[20–23]. To date, elongated vortices with C$_2$ symmetry are observed in the superconducting system exhibiting anisotropic electronic structure[24–26], confined geometries[27], heterostructures[28], and at step edges, where Abrikosov-Josephson vortices can emerge[29]. Beyond these findings, recent studies also highlight a strong interplay between charge density waves (CDWs) and superconductivity[30–32], particularly in quasi-two-dimensional transition metal dichalcogenides (TMD)[33–36] combining intricate effects of thickness and stacking fault with the physics of superconductivity. However, the influence of CDWs on vortex behavior and its potential for enabling vortex manipulation remains an open and compelling question.

In this work, we report the tunable bound states in the symmetry-breaking superconducting vortices via unidirectional CDWs in a TMD superconductor 1T″-NbTe$_2$ by utilizing scanning tunneling microscope/spectroscopy (STM/STS) at an ultralow temperature of 5 mK with an effective temperature of about 138 mK. We demonstrate that the 3×1×3 CDW order exhibits a three-dimensional character across single-unit-cell step edges. This stripe order coexists with superconductivity, which emerges below a critical temperature of 0.4 K. Under an applied magnetic field perpendicular to the sample surface below a critical magnetic field of 30 mT, we observe elliptical vortices whose long axes align with the 3×1 CDW stripes, indicating strong coupling between superconductivity and the underlying electronic modulation. The anisotropy of elongated vortex is estimated to be $\xi_{//} / \xi_{\perp} = 2.39 \pm 0.03$. Moreover, CDW domain boundaries induce distinct reorientations and deformations in the vortex shapes, effectively modulating vortex configurations across CDW nanodomains.

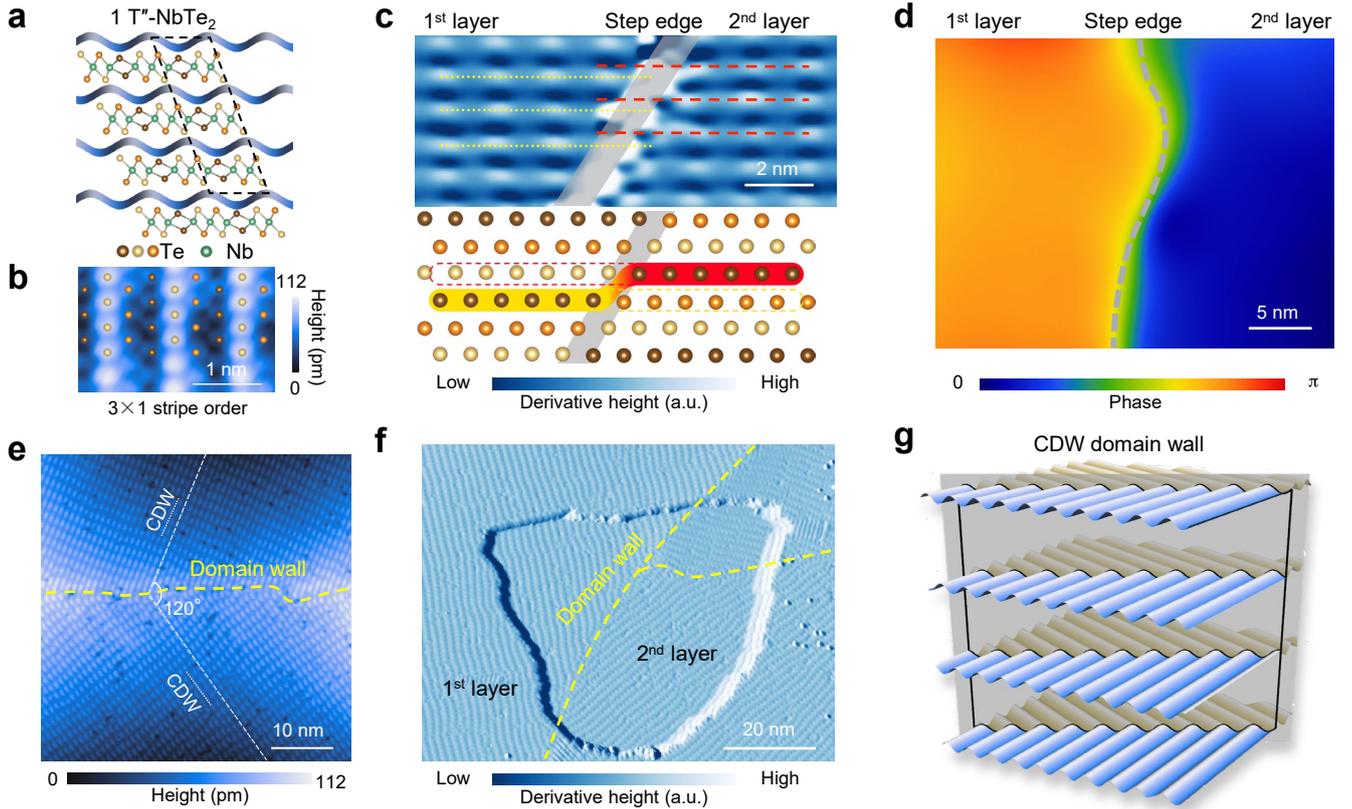

**Fig. 1. Three-dimensional unidirectional 3×1×3 CDW with the domain walls in 1T″-NbTe₂. a**, Schematic illustration of the layered crystal structure of 1T″-NbTe₂. The dashed lines mark one unit cell. **b**, Atomically resolved STM topographic image showing the 3×1 periodic atomic modulation at the tellurium-terminated surface ($V_s$ = -1 mV, $I_t$ = 1 nA). **c**, Differential STM image and schematic of a single-unit-cell step edge, showing the reveal lateral displacement of one-third the 3×1 CDW period between upper and lower layer. The yellow (red) dashed lines highlight the 3×1 CDW periodicity in the upper (lower) layer. ($V_s$ = -100 mV, $I_t$ = 100 pA). **d**, Local phase map of the 3×1 CDW obtained by 2D lock-in analysis, showing a $2\pi/3$ phase shift across the step edge (white solid lines). ($V_s$ = -100 mV, $I_t$ = 100 pA). **e**, STM topographic image, showing a domain wall (yellow dashes) separating two CDW domains with an orientation difference of 120° (white dashes). The white dotted lines highlight the direction of unidirectional CDW vector in each domain. ($V_s$ = -300 mV, $I_t$ = 50 pA). **f**, Differential STM topographic image with the contrast of both CDWs and domain walls enhanced, showing continuous CDW domain walls (yellow dashes) across the step edges. **g**, Schematic showing the 3D unidirectional CDW and a domain wall spanning over the entire crystal along the *c* axis.

We first demonstrate the formation of 3×1×3 stripe order in the bulk $NbTe_2$. Bulk $NbTe_2$ crystallizes in the 1T structure at high temperatures. Below a critical temperature of 530 K, it has been reported to undergo a structural transition to monoclinic 1T″ phase with a three-dimensional 3×1×3 periodic CDW order (Figure 1(a))[37–39]. The 1T″-$NbTe_2$ exhibits layered structure where each layer contains one layer of niobium (Nb) atoms sandwiched by two layers of tellurium (Te) atoms. The cleavage of 1T″-$NbTe_2$ crystal results in the Te-terminated surfaces, which show 3×1 periodic atomic modulation in the high-resolution STM images obtained at low temperature (Figure 1(b)). In addition to the 3×1 modulation, another 1×9/2 CDW[40] orthogonal to the 3×1 order is also observed in the topographic images of Te-terminated surface (Figure S1). The periodicity along *c* axis of the CDW order with phase twist is checked at a single-unit-cell step edge with a height of about 0.69 nm, which allows us to simultaneously view the stripe of the upper and lower Te-terminated surfaces side by side (Figure S2). The differential analysis of atomically-resolved topographic image reveals that the stripe order exhibits a relative phase shift across the single-unit-cell step edge (Figure 1(c)). By employing two-dimensional lock-in technique[41], the averaged phase difference between the adjacent layers is determined to be $2\pi/3$ (Figure 1(d), details see Methods and Figure S3), which demonstrates the formation of 3×1×3 stripe order at the atomic scale.

In addition to the 3×1 modulation at the single-domain surface region, multiple domains of CDWs with three orientations rotated by 120 degrees are observed in STM images of Te-terminated surfaces (Figure 1 (e)). It is worth nothing that the 1×9/2 CDW is orthogonal to the 3×1 CDW in each nanodomain. The observation of multiple unidirectional nanodomains with orientation difference of 120º may result from the spontaneous symmetry breaking after the CDW phase transition. Given that the 3×1 modulation is periodically stacked along the *c*-axis, a natural question arises: are the associated domain walls in STM images merely surface defects, or do they extend through the bulk along *c*-axis as two-dimensional structures? To address this, we analyze domain walls crossing a single-unit-cell step edge. We find that domain walls are not interrupted by the step edge, indicating that these domain walls are surfaces passing through the entire crystal instead of lines inside each surface (Figure 1(f), details see Figure S2). Therefore, the domain wall observed on the Te surface are two-dimensional wall connecting two 3×1×3 striped order in the bulk $NbTe_2$ (Figure 1(g)).

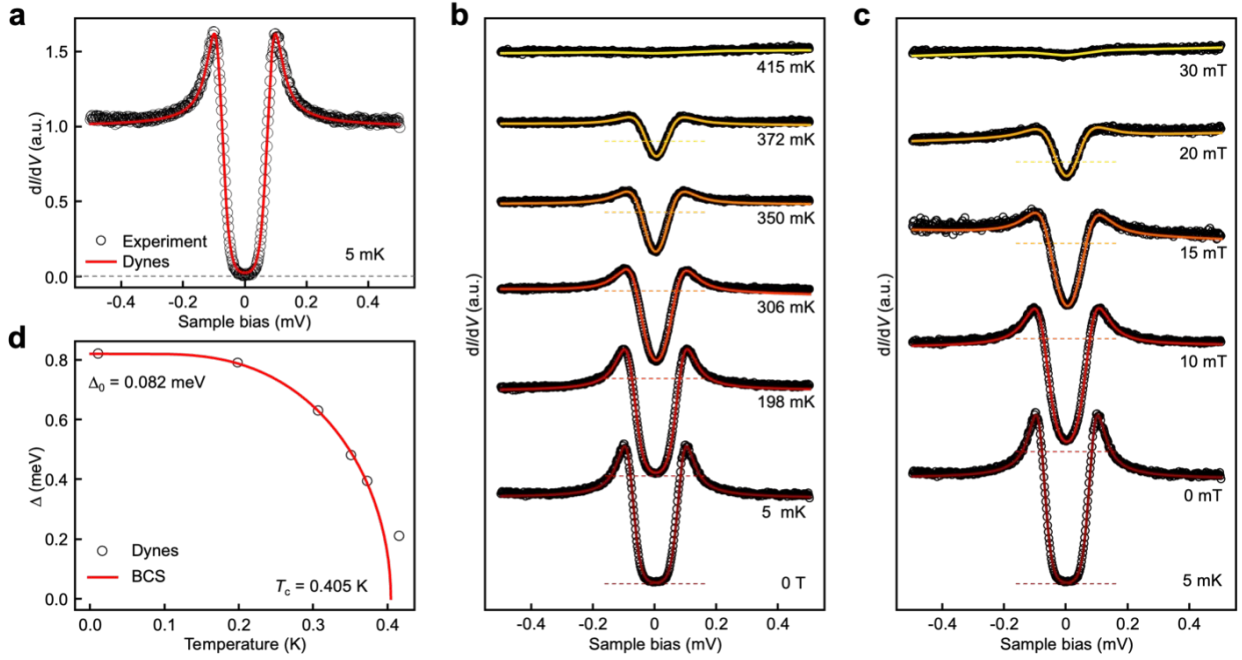

**Fig. 2. Temperature and magnetic dependent of d$I$/d$V$ spectra of 1T″-NbTe$_2$. a**, Typical d$I$/d$V$ spectrum (black circles) acquired under $T_{base}$ = 5 mK and $\mu H$ = 0 T, along with Dynes function description (red curve). **b**, Temperature-dependent d$I$/d$V$ spectra (black circles) and corresponding Dynes-function description (color curves) over the temperature ranging from 5 mK to 415 mK, showing the gradually suppression of the energy gap with increasing temperature. **c**, Out-of-plane magnetic field dependent d$I$/d$V$ spectra (black circles) and corresponding Dynes-function description (color curves) from 0 to 30 mT, showing the gradually suppression of the energy gap with increasing field. To ensure that the measured spectra are representative of the superconducting state at each magnetic field, all spectra were intentionally taken at the positions far away from the vortex cores in the zero-energy conductance maps. **d**, Temperature dependence of superconducting energy gap, showing the critical temperature of superconductivity $T_c$ ~ 0.4 K. Spectra in (b) and (c) are shifted vertically for clarification with dashed lines indicating zero-conductance. STS parameters: $V_s$ = 1 mV, $I_t$ = 1 nA, $V_{mod}$ = 5 μV.

We then investigate the density of states on the surface of 1T″-NbTe$_2$ in the superconducting state. The differential conductance (d$I$/d$V$) spectrum is proportional to the local density of states of the sample surface at low energies, which can provide key information on the superconducting gap at Fermi level. At a base temperature of 5 mK, which is far below the critical superconducting temperature ($T_c$), the typical d$I$/d$V$ spectrum shows a particle-hole symmetric energy gap (Figure 2(a)). Such energy gap is well described by the Dynes function with a gap magnitude Δ ≈ 0.082 meV (details see Figure S4). To further confirm the superconducting origin of the energy gap, evolution of d$I$/d$V$ spectra with temperature (Figure 2(b) and (d)) and out-of-plane magnetic field (Figure 2(c)) are investigated. As temperature increases, the gap amplitude is suppressed gradually and vanishes at a temperature of approximately 0.4 K (Figure 2(d)), which is comparable to the transport experiments[37]. Furthermore, the gap is suppressed progressively upon the rising of magnetic field along *c*-axis and fully vanished at a field around $H_c$ = 30 mT. We thus conclude that the observed energy gap is the superconducting gap that opens at $T_c \sim 0.4$ K on the surface of 1T″-NbTe$_2$.

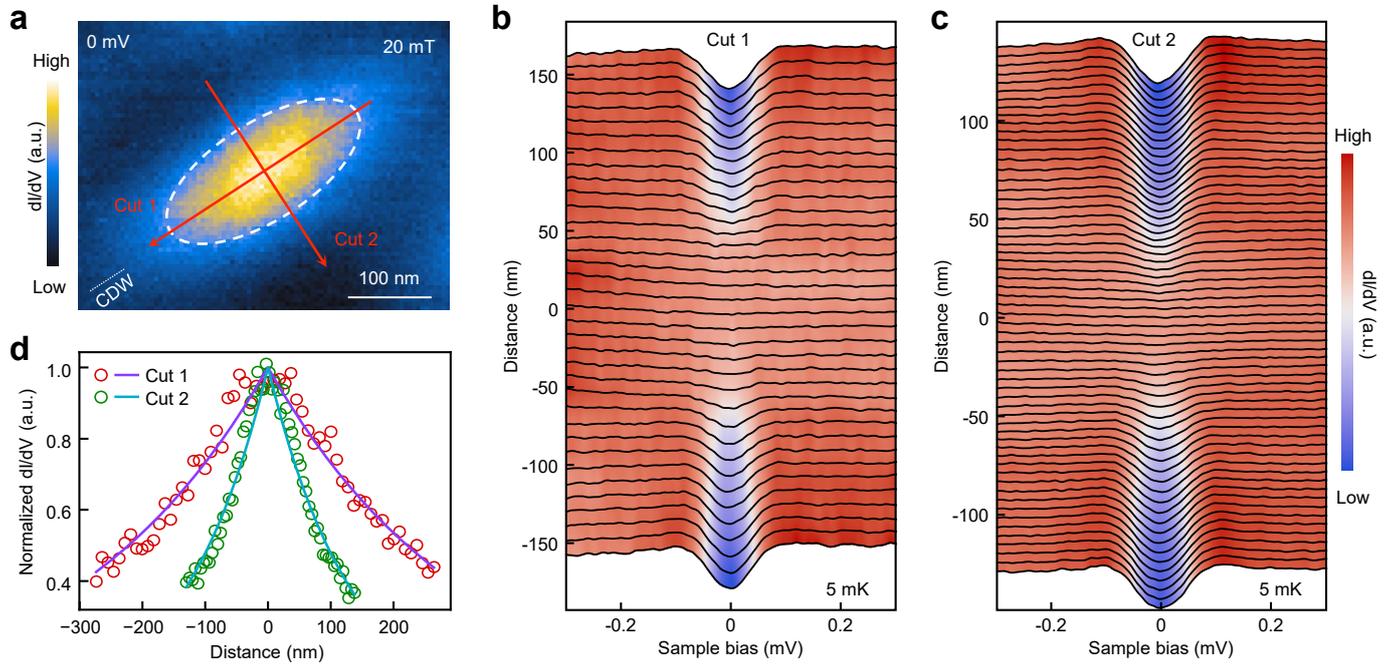

**Fig. 3. Elongated superconducting vortex aligned with the CDW stripe. a**, Zero-energy d$I$/d$V$ map over a single CDW domain under vertical magnetic field $\mu H$ = 20 mT, displaying an anisotropic superconducting vortex core elongated along the CDW stripe. A white dashed ellipse illustrates the oval shape of the vortex core as a visual guide. The white arrow dotted line is parallel to the 3×1 CDW, which is align with the elongated axis. **b,c**, Spatial evolution of d$I$/d$V$ spectra along Cut 1 (red arrow along short axis of the ellipse in (a)) and Cut 2 (red arrow along long axis of the ellipse in (a)) crossing the vortex core respectively. **d**, Exponential fitting (solid curves) of zero-energy conductance profiles along Cut 1 (red circles) and Cut 2 (green circles), showing the coherence lengths $\xi_1$ = 320.0 nm, $\xi_2$ = 135.5 nm with strong anisotropy $\xi_1/\xi_2 \sim$ 2.36 of the vortex core. $V_s$ = -10 mV, $I_t$ = 2 nA, $V_{mod}$ = 20 μV.

When the magnetic field is smaller than $H_c$ (e.g. 20 mT), the superconducting vortices are mapped out in the zero-energy conductance map, indicating that 1T″-NbTe$_2$ is a type II superconductor[2]. Strikingly distinct from the isotropic vortex in most superconductors[15], the single vortex in NbTe$_2$ exhibits anisotropic shape with two-fold symmetry (Figure 3(a)). In addition, the long (short) axis of elongated vortex aligns with the 3×1 (1×9/2) stripe order at Te surface, indicating a strong coupling between the superconducting vortex flux and the unidirectional CDWs. As the 3×1×3 modulation has been established as a three-dimensional CDW in bulk 1T″-NbTe$_2$, while the 1×9/2 CDW is likely a surface charge order, we mainly focus on the correlations between the bulk 3×1×3 CDW and the bulk superconducting vortices in the following discussion. From d$I$/d$V$ spectra taken along the long axis (Cut 1, Figure 3(b)) and short axis (Cut 2, Figure 3(c)) of the elongated vortex, the evolution of zero-energy conductance to the distance from the vortex center is fitted with exponential decay (details see Methods), giving the coherence length ($\xi$) along two orthogonal directions and the anisotropy ratio $\xi_1/\xi_2$ ~2.39 ± 0.03 (Figure 3(d) and Figure S10). Additionally, flat d$I$/d$V$ spectrum is observed at the center of the vortex, distinct from the conventional Caroli-de Gennes-Matricon (CdGM) vortex bound states, typically presenting as a prominent zero-bias peak, in superconductors not being in the quantum limit[15]. The absence of a zero-bias peak might suggest that the superconductor is at a dirty limit, where quasiparticle mean free path is lower than the coherence length[42] or exhibits a multiband structure[43]. Scattering of defects can also smear out the zero-bias peaks, which is unlikely given the high quality of the cleavage surface in our experiments.

Anisotropic vortex is usually related with Fermi surface anisotropy or superconducting order parameter anisotropy, where the symmetry of the vortex reflects the symmetry of the underling electronic structure[44,45]. The Fermi surface of undistorted NbTe$_2$ has six-fold symmetry with possible nesting vector $q = 1/3$ $a$. A recent STM experiment suggested that the periodic lattice distortions (PLDs) result from the formation of multiband CDWs through Fermi-surface nesting[40]. Moreover, the development of the CDWs significantly modifies the shape and symmetry of the Fermi surface. It is also worth noting that lattice distortions, such as PLDs[46,47] and nematic transitions[48–50], have been recognized to strongly affect both the electronic structure and the superconducting gap. Therefore, it is reasonable to deduce that the formation of unidirectional CDWs/PLDs has led to further reconstruction of the electronic structure, resulting in strong anisotropic Fermi surface and superconducting order parameter.

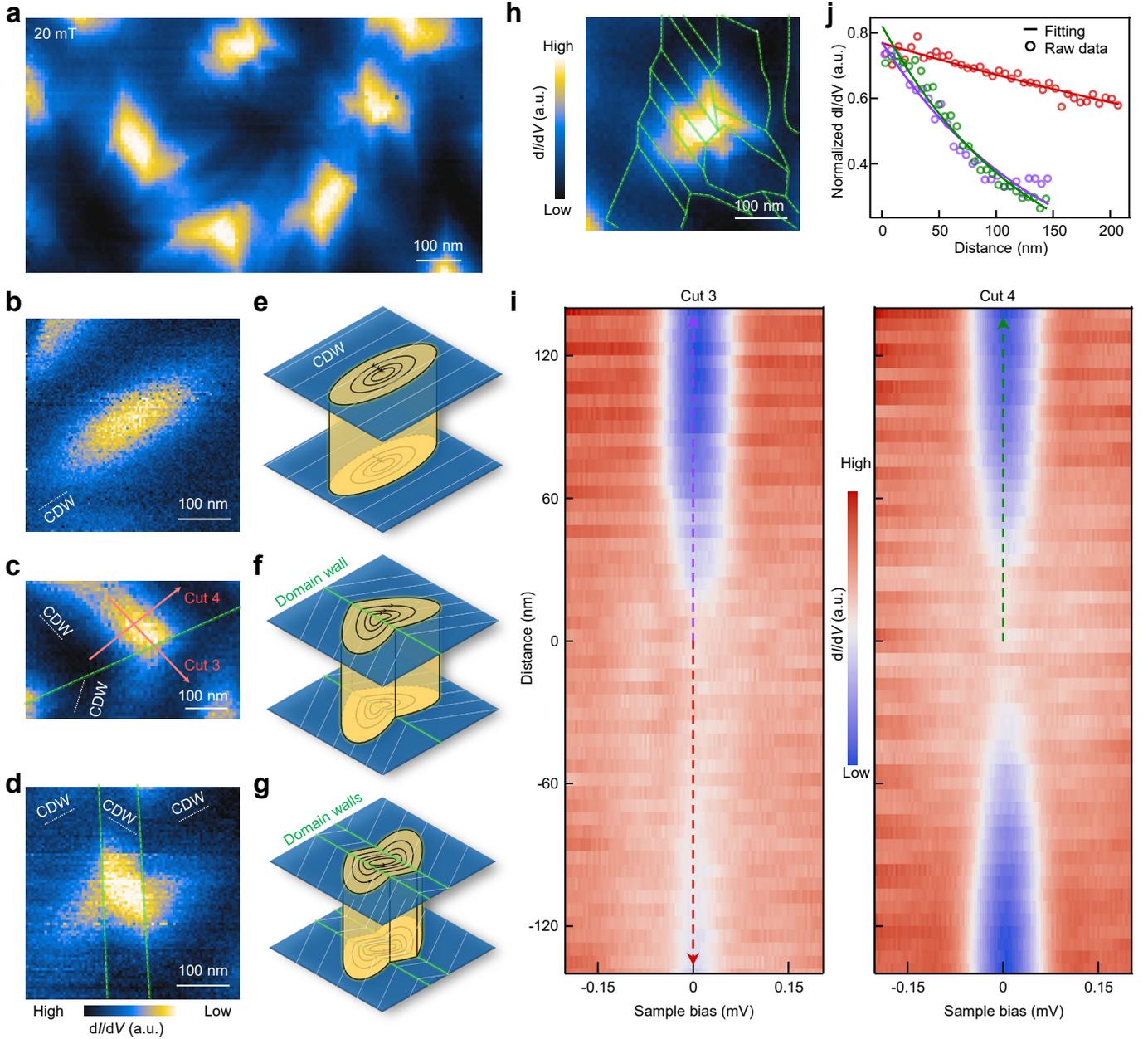

**Fig. 4. Tuning of the quantum flux and bound states in the symmetry-breaking vortex core by unidirectional CDW nanodomains. a,** Zero-energy map under a magnetic field of 0.02 T, showing the vortex lattice. **b-d,** Zero-energy conductance maps of surface region with 1 (b), 2 (c), 3 (d) CDW nanodomains, respectively, showing that the orientation of vortex cores rotates to match the direction of CDW stripes (white lines parallel to 3×1 CDW stripe in each nanodomains). **e-g,** Schematic models of the vortex core morphologies crossing 0, 1 and 2 CDW domain walls corresponding to the vortex map in (b-d), showing the modification of vortex quantum flux by the two-dimensional CDW domain walls. **h,** Zero-bias conductance maps of a surface region with multiple CDW domains, showing a complex geometry of vortex core. **i,** Spatial evolution of d$I$/d$V$ spectra along Cut 3 crossing a domain wall (left) and Cut 4 within single domain (right) in (c), revealing asymmetric decay on two sides of domain wall (left) and conventional symmetric decay similar to an isolated vortex within a single domain. **j,** Zero-bias conductance profiles versus distance from the vortex core center (circles), with exponential fits (solid lines), showing the distinct coherence length across the domain wall. $V_s$ = -10 mV, $I_t$ = 2 nA, $V_{mod}$ = 20 μV.

The geometry of vortex structures is strongly modulated by the nanoscale domains of CDW order. Low-temperature cleaving of 1T″-NbTe$_2$ produces multiple CDW nanodomains, typically with lateral dimensions of tens of nanometers. On such a surface, application of external out-of-plane magnetic field leads to a disordered vortex arrangement, in stark contrast to the regular hexagonal or square lattice in conventional type-II superconductors (Figure 4(a) and Figure S5). Among the disordered arrangement, individual vortices traversing multiple CDW domains exhibits complex, nontrivial shapes (see Method for details). When the size of CDW domain exceeds the coherence length of superconductivity, the vortex is elongated with the long axis aligned to the unidirectional CDW (Figure 4(b) and (e)). However, in most cases, the average 3×1×3 CDW domain size is smaller than the coherence length, causing single vortices to span several nanodomains and cross multiple domain walls. When a single vortex cross two neighboring CDW nanodomains, it adopts a characteristic V-shaped geometry, with the orientation of each segment aligned with the respective stripe order in each domain (Figure 4(c) and (f)). In cases where a vortex traverses three nanodomains separated by parallel domain walls, it takes on an N-shaped configuration, with the long axis changing direction twice by 120°, consistent with the angular relationship between adjacent domains (Figure 4(d) and (g)). As the number of intersected domains increases, the vortex shape becomes increasingly irregular (Figure 4(h)). These results demonstrate that the real-space configuration of quantized magnetic flux can be finely tailored by engineering the arrangement of CDW nanodomains under an applied magnetic field.

Beyond the vortex morphology, the local electronic structure, particularly the vortex bound states, is also modulated by the CDW domain structure. Inside the vortex across two CDW nanodomains, the d$I$/d$V$ spectra crossing the domain wall (Cut 3 in Figure 4(c)) show that the CdGM states evolute asymmetrically in space (left panel of Figure 4(i)), in contrast to the symmetric dependence (right panel of Figure 4(i)) inside a single domain (Cut 4 in Figure 4(c)). To compare the coherence length on both sides of the CDW domain wall, zero-energy conductance as the function of distance from the vortex core is plotted together in Figure 4(j). On the line cutting through the domain wall (Cut 3), contrasting decay lengths exists, matching Cut 4 on one side but differing significantly on the other side, implying a sudden turn of the vortex orientation. For vortices spanning multiple CDW nanodomains, the vortex bound state become increasingly complex, potentially due to enhanced anisotropy of the Fermi surface and order parameter[45]. The observation of elongated and symmetry-breaking vortices, controlled by nanoscale CDW textures, is exceptionally rare in type-II superconductors. To date, no prior study has reported on vortex bound states across multiple anisotropic-order domain boundaries. Our findings open new possibilities for exploring vortex physics in superconductors with multiple CDW nanodomains, and suggest a novel platform for engineered vortex manipulation at the nanoscale.

In summary, our study reveals that stripe-ordered 3×1×3 CDW in 1T″-NbTe$_2$ strongly influence the spatial symmetry and dynamics of superconducting vortices. The emergence of elongated vortices aligned with the stripe order and their reorientation across domain boundaries demonstrate a direct coupling between superconducting and electronic orders. This coupling enables a new degree of control over vortex configurations by tuning the underlying CDW landscape. Such control offers a new platform for engineering vortex-bound states, with potential applications in topological quantum computation and reconfigurable superconducting devices.

## METHODS

**Single-Crystal Growth of NbTe$_2$.** The single crystals of NbTe$_2$ (Space group: C2/m, No. 12) were synthesized by chemical vapor transport (CVT) approach. Amounts of Nb (Alfa, 99.999%) and Te (Alfa, 99.999%) in a molar ratio of Nb:Te = 1:2, along with I$_2$ as the transport agent, were thoroughly mixed and sealed in a quartz tube with a high vacuum. Subsequently, the growth of NbTe$_2$ single crystals was carried out in a two-zone tube furnace with a temperature gradient from 850 °C (source zone) to 750 °C (growth zone). After about 7 days, large-size (~5 mm) and high-quality NbTe$_2$ single crystals were collected at the low-temperature part.

**Scanning tunneling microscopy/spectroscopy.** Experiments were performed in an ultrahigh vacuum (1×10$^{-10}$ mbar) ultra-low temperature STM system equipped with external magnetic field perpendicular to the sample surface. The NbTe$_2$ samples used in the STM/STS experiments were cleaved at room temperature (300 K) in an ultrahigh vacuum chamber, and were then *in-situ* transferred to the STM scanner and cooled down to about 6 K. The lowest base temperature is 5 mK with an electronic temperature of 138 mK. At such electronic temperature, the W tip is not superconducting. All the scanning parameters (setpoint voltage $V_s$ and tunneling current $I_t$) of the STM topographic images are listed in the figure captions. The d$I$/d$V$ spectra were acquired by a standard lock-in amplifier at a modulation frequency of 877.1 Hz, while the bias modulation amplitudes ($V_{mod}$) are listed in the figure captions. Non-magnetic tungsten tips were fabricated via electrochemical etching and calibrated on a clean Au(111) surface prepared by repeated cycles of sputtering with argon ions and annealing at 500 °C.

**2D lock-in technique.** To retrieve the phase information of the CDW, we used a 2D lock-in technique adapted from Lawler-Fujita drift correction[51]. The CDW observed in STM topography $T(\mathbf{r})$ can be described by

$$T(\mathbf{r}) = A(\mathbf{r})e^{i\phi(\mathbf{r})}\left(e^{i\mathbf{q}\cdot\mathbf{r}} + e^{-i\mathbf{q}\cdot\mathbf{r}}\right),$$

where $A(\mathbf{r})$ represents the spatial dependent amplitude, $\phi(\mathbf{r})$ is the local phase, and $\mathbf{q}$ is the reciprocal vector of a CDW. Following the lock-in procedure, the topography $T(\mathbf{r})$ is multiplied with a 2D reference signal $e^{-i\mathbf{q}\cdot\mathbf{r}}$, followed by the employment of a Gaussian low pass filter

$$f(\mathbf{r}) = \left[T(\mathbf{r}) \cdot e^{-i\mathbf{q}\cdot\mathbf{r}}\right] * e^{-\frac{\mathbf{r}^2}{2\sigma^2}} \approx A(\mathbf{r})e^{i\phi(\mathbf{r})}.$$

The local phase $\phi(\mathbf{r})$ of the CDW can be directly acquired from the output of the 2D lock-in technique $f(\mathbf{r})$. Figure S3 shows the process retrieving the phase information displayed in Figure 1(d), using the 2D lock-in technique. The reciprocal vector of the CDW is marked by red circles in Figure S3(c). The amplitude $A(\mathbf{r})$ and local phase $\phi(\mathbf{r})$ acquired from $f(\mathbf{r})$ described in the upper equation is presented in Figure S3(e) and (f). Notably, a slow varying background exists in the spatial dependence (Figure S3(f) and (g)), which may be contributed by piezoelectric nonlinearity, thermal drift, and hysteretic effects of the STM scanner. A linear fitting is used to remove this background. The final result of the 2D lock-in process is shown in Figure S3(h) from which the phase jump across the step edge is quantitatively measured to be about $2\pi/3$.

**Measuring the coherence length from vortices.** The coherence length of superconductivity is measured by fit the decay of zero-bias conductance with an exponential function. Since a flat dI/dV spectrum is observed in our measurements, the normalized differential conductance at the vortex center should be 1. As a result, each dI/dV spectrum is normalized to the conductance outside the superconducting gap and then fitted with the following exponential decay

$$g_{norm}(r) = (1 - g_{norm}(\infty))e^{-\frac{|r-r_0|}{\xi}} + g_{norm}(\infty)$$

where $g_{norm}(\infty)$ is the normalized dI/dV far away from the vortex center, $r_0$ is the location of vortex center and $\xi$ is the coherence length.

**Characterizations of CDW nanodomain and vortices.** The positions of CDW domain walls are determined by the differential topographic image where the intensity changes with the CDW orientation (Figure S6). At zero field, the superconducting gap is not suppressed at the CDW domain wall (Figure S7). The spontaneously obtained zero-energy map under a magnetic field shows the structure of vortices. In addition, the locations of vortices are controlled by switching the magnetic field (Figure S8), which excludes the pining effect of CDW domain walls. The observations of CDW domain wall tunable vortices are repeatable in total four $NbTe_2$ samples which are cleaved at room temperature (Figure S9).

## ACKNOWLEDGMENTS
The work is supported by grants from the National Natural Science Foundation of China (62488201), the National Key Research and Development Projects of China (2022YFA1204100), the CAS Project for Young Scientists in Basic Research (YSBR-003) and the Innovation Program of Quantum Science and Technology (2021ZD0302700).